\documentstyle[preprint,eqsecnum,aps,epsf,psfig]{revtex}
\newif\iftightenlines\tightenlinesfalse
\tightenlines\tightenlinestrue
\begin{document}
\draft
\preprint{\vbox{\hbox{CERN-TH/97-283}
\hbox{AZPH-TH/97-12}
\hbox{October 1997}}}
\title{Secondary Decays in Atmospheric Charm Contributions
to the Flux of Muons and Muon Neutrinos}
\author{L. Pasquali$^1$, M. H. Reno$^{1,2}$ and I. Sarcevic$^3$}
\address{
$^1$Department of Physics and Astronomy, University of Iowa, Iowa City,
Iowa 52242\\
$^2$CERN-TH, CH-1211 Geneva 23, Switzerland\\
$^3$Department of Physics, University of Arizona, Tucson, Arizona
85721}

\maketitle

\begin{abstract}
We present a calculation of the fluxes
of muons and muon neutrinos
from the decays of pions and kaons that are themselves the
decay products of charmed particles produced in
the atmosphere by cosmic ray-air collisions. 
Using the perturbative cross section for charm
production, these lepton fluxes are two to three orders of
magnitude smaller than the fluxes from the decays of pions and kaons
directly produced in cosmic ray-air collisions.
Intrinsic charm models do not significantly alter our conclusions, nor
do models with a charm cross section enhanced in the region above
an incident cosmic ray energy of 1 TeV.

\end{abstract}

\section{INTRODUCTION}

The fluxes of atmospheric
neutrinos and muons with energies larger than 100 GeV, which result
from cosmic ray interactions with air molecules
in the atmosphere\cite{volkova,tig},
are measured in large underground detectors\cite{lud}.
The fluxes are interesting for a variety of reasons, including the fact that
the leptons are the decay products of particles
produced in collisions with center of mass energies that
may be extremely high.
Cosmic rays, comprised primarily of protons, have been measured with energies
up to $E\sim 10^{20}$ eV\cite{fe}, many orders of magnitude beyond the
energies accessible in accelerator laboratories.
In addition, atmospheric neutrinos and muons are a background 
for galactic and extra-galactic sources of neutrinos\cite{gqrs}.

Atmospheric lepton fluxes come from two main sources.
The so-called ``conventional'' atmospheric fluxes of muons and neutrinos
come from
hadronic production of $K^\pm$'s, $K_L$ and $\pi^\pm$'s in cosmic ray-air
collisions, followed by their leptonic, and in the case of kaons,
semileptonic, decays. At the energies considered here, $E>10^2$ GeV,
the magnitudes of the conventional lepton fluxes
are governed by the lifetimes of the mesons. Because of time dilation,
the decay lengths of these mesons are much larger than the depth
of the atmosphere. A small fraction decay to leptons, but since they are
copiously produced, the conventional flux dominates the lepton flux at
the lower range of energies considered here. Given an incident
cosmic ray flux at the top of the atmosphere of
$\phi_{CR}\sim E^{-\alpha}$, the
flux of leptons from $K$ and $\pi$ decays has the
energy dependence $\phi_{\nu,\mu}\sim E^{-\alpha-1}$. Values of $\alpha$ 
are measured with $\alpha\sim 2.7-3$\cite{crs}. 

Charmed particle production and decay make important contributions to the
lepton fluxes at high energies. Essentially all of the 
charmed particles decay in the atmosphere
since $\tau\sim 10^{-12}$ s, so these contributions are said to contribute to
the ``prompt'' atmospheric flux. The energy behavior of the prompt fluxes
is one power higher than the conventional fluxes for $E<10^7$ GeV: 
$\phi_{\nu,\mu}\sim E^{-\alpha}$. Even though charm production is
suppressed relative to pion and kaon production, the energy behavior of 
the prompt flux means that the flux of leptons from charm decay dominates
at high energies.
In the recent work of Thunman, Ingelman
and Gondolo\cite{tig}, labeled here as TIG, a Monte Carlo model of
atmospheric lepton production based on PYTHIA\cite{pythia} 
was used to calculate
the contribution of charmed particle semileptonic decays to the atmospheric
flux.

Along with the semileptonic decays of charmed particles, in principle,
secondary decays such as $D\rightarrow K\rightarrow \mu\nu_\mu$
also contribute to the atmospheric lepton fluxes.
Essentially all charmed meson decays have at least one kaon or pion in
the final state, and the charged kaons and pions decay mainly to
muons and neutrinos. It is the fluxes of these secondary muons and
neutrinos from charm decay, and the more general topic of the
uncertainties in the theoretical predictions for the
atmospheric lepton fluxes from charm decay, 
that are the topics of this paper. We
show that
using the perturbative cross section for charm production, the secondary
lepton fluxes are significant for $E<10^3$ GeV relative to the
prompt fluxes, however they are small compared to the conventional fluxes.
Even for strongly
enhanced charm production, as long as the cross section is 
consistent with experimental measurements,
the secondary lepton
flux is negligible compared
to the conventional flux.

In Section 2, we outline the standard
calculational procedure, an approximate
analytic method. Details of this procedure are outlined in
Refs. \cite{tig,lipari,gaisser}.
We list our numerical inputs in Section 3
and present our results in Section 4.
We follow with a summary in Section 5.

\section{Calculational Procedure}

The calculations presented here are based on the cascade equations 
for baryon, meson and lepton 
fluxes\cite{tig,lipari,gaisser}.
The cascade equations describe the production of particle $j$ through
interactions and decays of particle $k$, and the loss of particle
$j$ through absorption and its own decay.
The energy dependent 
flux of particles  of type $j$, $\phi_j(E,X)$, is governed
in general by the equation
\begin{equation}
{{\rm d}\phi_j\over {\rm d}X} = -{\phi_j\over \lambda_j}
-{\phi_j\over \lambda^{(dec)}_j}+\sum_k S(k\rightarrow j)
\end{equation}
where $X$ is the column depth (slant depth)
in the atmosphere measured from the top of the atmosphere, 
$\lambda_j$ is the
hadronic interaction length and $\lambda^{(dec)}_j\sim
\gamma c \tau_j$ is the decay length,
all converted to units of g/cm$^2$. 
The column depth is dependent on the incident
angle. For vertical fluxes evaluated at sea level, 
the column depth is $X=X_0=1300$ g/cm$^2$.  The quantity $S(k\rightarrow j)$
describes the source of particles of type $j$ from interactions or
decays of
particles of type $k$. For neutrinos and muons
at energies above 100 GeV, since the interaction
and decay lengths are essentially infinite, only the source terms 
appear on the right hand side of equation (2.1).
The source terms have the form:
\begin{equation}
S(k\rightarrow j)=\langle N_j\rangle \int_E^\infty {\rm d}\,E_k
{\phi_k(E_k,X)\over \lambda_k(E_k)}
 {{\rm d}n_{k\rightarrow j}(E;E_k)
\over {\rm d}E}\ , 
\end{equation}
where $\lambda_k(E_k)$ is an interaction length or decay 
length, depending on the source. The quantity d$n_{k\rightarrow j}$/d$E$
describes the energy distribution of the produced particle $j$, and
$\langle N_j\rangle$ is the particle multiplicity.
For decays, the value of 
$\langle N_j\rangle$ is unity, except for pions.
For the production of charm, $\langle N_c\rangle = 2$.

In principle, the cascade equations are a coupled set of equations.
We make several simplifying assumptions so that we can solve the
equations analytically. 
First, we decouple the 
meson equations,
with the exception of charm decay to  pions and kaons. 
In principle, there are interaction terms such as 
$S(\pi\rightarrow K)$, however, these are small compared to 
$S(p\rightarrow K)$, so we neglect the former.
With this assumption, we can solve for 
conventional $\pi$ and $K$ fluxes separately,
then evaluate their contributions separately to the neutrino and muon
fluxes. The same holds true for charm production. We follow the produced
charm in their decays to leptons, pions and kaons.

To further simplify the calculation, we represent the cosmic
ray flux by the proton flux. The cosmic ray composition is dominantly
protons\cite{nf}, so the treatment of cosmic
rays as protons is reasonable. This approximation was used by TIG in
Ref. \cite{tig}. A feature that allows us to approximately solve
the cascade equations is to use a factorization of the energy
dependence in the fluxes. By writing 
\begin{equation}
\phi_j(E,X)\equiv f_j(E,X)E^{-\beta_j}\ ,
\end{equation}
and assuming that $f_j(E,X)$ is weakly dependent on energy,
approximate solutions to the cascade equations are straightforward and lead
to the power law energy behavior described in the introduction.
Thunman, Ingelman and Gondolo have demonstrated this is a good approximation
by comparing the approximate solutions to Monte Carlo results \cite{tig}.
This is especially useful because the source terms factorize to
\begin{equation}
S(k\rightarrow j) = {f_k(X)E^{-\beta_k}\over \lambda_k(E)}
Z_{kj}(\beta_k,E )
\end{equation}
where $Z_{kj}(\beta_k,E)$ depends only weakly on energy.
The calculation of the $Z$-moments is the essential ingredient in
addition to $\lambda$ to describe the lepton fluxes. The interaction moments
include information about multiplicity, 
the magnitude and energy dependence of the relevant cross section 
and the energy distribution of the emerging particle. 
To simplify notation, we omit the energy argument  in $Z$'s, although
we do include energy dependent $Z$'s in our evaluation of the
fluxes.

With the assumptions above, 
the solutions for the hadron fluxes are straightforward.
For cosmic ray protons, $\lambda^{(dec)}_p$ is infinite, and
the solution for the flux is
\begin{equation}
\phi_p(X,E)=f_p(X)E^{-\alpha}=f_p(0) e^{-X/ \Lambda_p}E^{-\alpha} 
\end{equation}
for $\Lambda_k\equiv \lambda_k/(1-Z_{kk})$.
For hadrons other than the proton, the solutions to the factorized cascade
equations have two different forms, depending on whether one is in
the high or low energy regime. The energy regime is determined
by whether or not 
the decay length of the particle is large compared to the column depth.
The cascade equations are solved approximately in these two regimes,
then interpolation between high and low is done with
\begin{equation}
\phi_j={\phi_j^{low}\phi_j^{high}\over
\phi_j^{low}+\phi_j^{high}}\ .
\end{equation}
Explicit solutions appear, for example, in Ref. \cite{lipari}.
The fluxes from each source are evaluated this way, then summed to
get the total flux.

The lepton fluxes come from hadron decays, where the hadron fluxes
are approximated via Eq. (2.6). In the high and low energy regimes,
the approximate lepton fluxes can be written in simple form. For
definiteness, we write the flux for neutrinos. The flux for
muons is in exactly the same form. At high energies, the (direct) flux
from source terms  $p\rightarrow j\rightarrow$leptons is
\begin{equation}
\phi_\nu^{j,high}(X,E) =  f_\nu^{j,high}(E,X)E^{-\alpha-1}=
{ Z_{pj}(\alpha )Z_{j\nu}(\beta_j)\over 1-Z_{pp}(\alpha) }
{\ln (\Lambda_j/\Lambda_p) \over 1-\Lambda_p/\Lambda_j }
{ f_p(0)\over \lambda_j^{(dec)}\rho(X)} E^{-\alpha}\ ,
\end{equation}
where $\rho(X)$ is the density of air at column depth
$X$. The energy dependence of $\lambda_j^{(dec)}$ makes the flux
of leptons from high energy hadron decays suppressed by one power
of energy relative to the proton flux.
For low energies, the hadron flux itself is proportional to 
$\lambda_j^{(dec)}$ and 
the decay length cancels. The energy dependence of the lepton
flux
tracks that of the primary cosmic ray proton flux.
From the direct decays of low energy hadrons $j$,
\begin{equation}
\phi_\nu^{j,low}(X,E) =  f_\nu^{j,low}(E,X)E^{-\alpha} = {Z_{pj}(\alpha)
Z_{j\nu}(\beta_j)
\over 1-Z_{pp}(\alpha)} f_p(0)E^{-\alpha}\ ,
\end{equation}

As indicated earlier, pions and kaons have relatively long lifetimes,
so for $E>10^2$ GeV, they are ``high energy'' hadrons
resulting in a lepton flux proportional to $E^{-\alpha-1}$. Charmed hadrons,
for most of the energy range $E\sim 10^2-10^8$ GeV, are low energy
particles, so the resulting lepton flux has the low energy form
proportional to $E^{-\alpha}$.

The extensions of these equations for secondary leptons 
from $p\rightarrow $charm$\rightarrow j\rightarrow$leptons is straightforward.
The substitution 
\begin{equation}
Z_{pj}(\alpha)\rightarrow Z_{pk_c}(\alpha)
Z_{k_cj}(\beta_{k_c})
\end{equation} 
is required for hadrons $j$
that come from charmed particle $(k_c)$ decays. 
The calculation of the moments for direct processes have be done with
a variety of assumptions \cite{lipari,tig}. 
It remains to indicate the inputs for interaction
lengths and $Z$-moments and to describe our calculation of 
$Z_{k_cj}(\beta_{k_c})$.

\section{Inputs}

Eqs. (2.7) and (2.8) govern the lepton fluxes, with the additional
substitution of Eq. (2.9) for the secondary fluxes.
We use TIG parameters where they are available \cite{tig}. TIG have
shown that for conventional charm production, $D^\pm$, $D^0$ and 
$\bar{D}^0$ decays dominate the sources of prompt leptons for most
of the energies between $10^2-10^8$ GeV. Consequently, we consider 
charm contributions only via $D^\pm$, $D^0$ and 
$\bar{D}^0$ decays. Henceforth, we will only refer to particles, however,
all of our calculations include both particles and antiparticles.
The inputs fall into three categories: the incident proton flux
(we assume that cosmic rays are protons), interaction $Z$-moments and
interaction lengths, and decay $Z$-moments and decay lengths.
 
The incident flux of protons at the top of the atmosphere, following TIG, 
is taken to be isotropic, with
\begin{eqnarray}
\phi_p(E,X=0) [{\rm cm^{-2}s^{-1}sr^{-1}GeV^{-1}}]\  =\ &
1.7\ (E/{\rm GeV})^{-2.7}\quad E<E_0\\ \nonumber
& 174\ (E/{\rm GeV})^{-3}\quad E\ge E_0\ ,
\end{eqnarray}
with $E_0=5\cdot 10^6$ GeV. 

The interaction lengths for the protons, pions
and kaons require the total scattering cross sections with isoscalar
nucleons. We use the Donnachie-Landshoff parameterization\cite{dl}
\begin{equation}
\sigma_{tot} = Xs^\epsilon +Ys^{-\eta}
\end{equation}
and the values of the parameters from the Particle Data Book\cite{pdg}
for incident protons, $\pi^+$ and $K^+$. We use the $K^+$ cross sections
for $K_L$, $D^0$ and $D^+$ as well. The cross sections are then rescaled
with a constant factor of $A^{2/3}$, where $A=14.5$ is the average atomic
number of air nuclear targets. 
Interaction $Z$-moments have been presented as a function of energy
in Ref. \cite{tig}  for nucleons, pions, kaons and $D$ mesons.
We use these $Z$-moments to evaluate Eqs. (2.7) and (2.8).

For particle lifetimes and branching fractions, we use the
Particle Data Book values.
The two-body decay moments for pion and kaon decays are straightforward
to calculate. Formulae appear in Ref. \cite{lipari}, and numerical
values, depending on the energy behavior of the decaying mesons,
appear in Refs. \cite{lipari} and \cite{tig}. For semileptonic and 
non-leptonic decays, the calculations are less obvious.
These must be done for the decay moments for charm to pions and kaons,
muons and neutrinos.
An accurate evaluation of $Z_{k_cj}(\beta_c)$ is difficult because
the hadronic decay distributions cannot be calculated analytically.
Data exist on distributions in charmed particle decays, for example,
the $K^\pm$ and $K^0$ momentum spectra for $D^0$, $D^+$ and $D_s^+$, where the 
$D$'s are all produced approximately at rest\cite{miii}. The data are
not immediately translatable into decay $Z$-moments because one
must boost to the frame where the kinetic energies of the $D$'s are
large.

For the simplest case of semileptonic
charm decays, TIG distribute lepton momenta according to a weak matrix
element
\begin{equation}
\mid {\cal M}\mid^2 = (p_D\cdot p_\mu)(p_\nu\cdot p_h)
\end{equation}
where $h$ is the final state hadron system.
They find that the decay moments to muons and neutrinos are equal to
within 15\%\ for $D$ mesons, and that the moments decrease by a factor
of $\sim 3$ as the $D$'s make the transition from the low energy
to high energy regime.

The matrix element in Eq. (3.3)
does not account for form factor modifications
of the hadron weak vertex.  
To look at the sensitivity of the decay moment to the
energy distribution, we have 
calculated the decay $Z_{D^\pm\nu}$ using three-body phase space
to dictate the energy distribution of the neutrino. Our results are shown
in Fig. 1 by the points. The energy dependence comes from the
changing energy behavior of the $D^\pm$ flux.
The upper dashed line is the TIG result
with $\beta_D=1.7$, appropriate for low energy $D$'s, for example, at 
$E=10^2$ GeV. The lower dashed line is with $\beta_D = 3$, appropriate for
high energy $D$'s, namely at $10^8$ GeV. Over the range of
energies between $10^2-10^8$ GeV, 
there is a factor of $\sim 2-2.4$ between the phase space
result and the relevant TIG numbers. This factor of $2-2.4$
gives a qualitative
estimate of the uncertainty in the decay moments. 
We use the TIG semileptonic decay moments,
interpolating between $10^2$ and $10^8$ GeV as a function of
$\ln (E/{\rm GeV})$ using a constant (negative) slope.

\vskip 0.5in
\centerline{\psfig{figure=zdpmnu.ps,height=3in,angle=270}}
\vskip 0.3in
\noindent
Fig. 1.
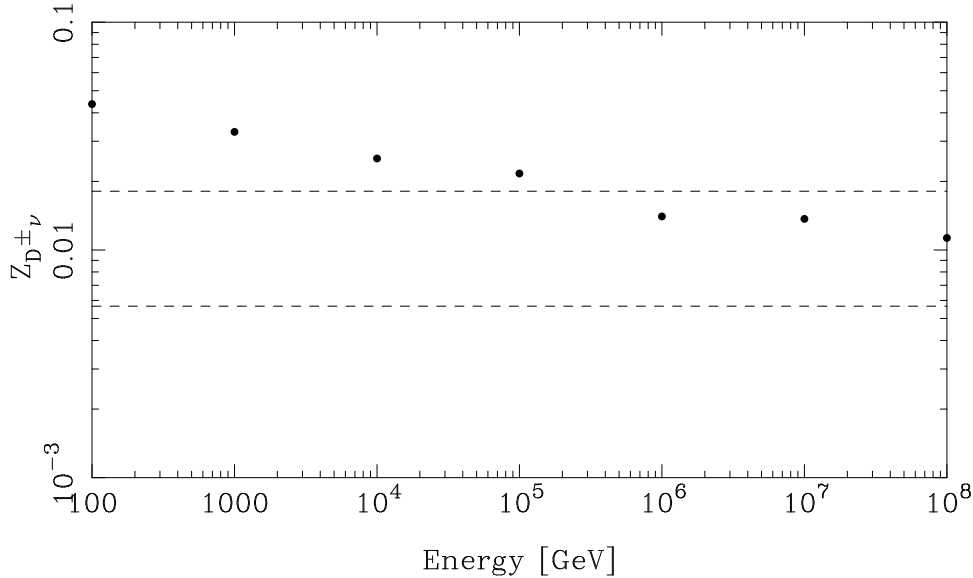
The points indicate the decay moment $Z_{D^\pm\nu}$ evaluated
using three-body phase space to distribute the neutrino energy.
The upper dashed line is the TIG result using $\beta_D=1.7$.
The lower dashed line has $\beta_D=3$ \cite{tig}.
\vskip 0.3in
 
Our procedure
for the hadronic decay moments of the $D$'s is to use the same
energy dependence and magnitude as the semileptonic $D$ moments
as a function of energy, with a correction
to account for the different 
branching fractions and particle multiplicities of the hadrons. 
For example,
\begin{equation}
Z_{D^\pm K^\pm} = Z_{D^\pm \nu}\cdot{\langle N_{K^\pm} 
\rangle \, B(D^\pm\rightarrow
K^\pm+{\rm anything})\over B(D^\pm\rightarrow \nu) }
\end{equation}

The $Z_{k_cj}(\beta_c=1.7)$-moments and the 
branching fractions and multiplicities 
used are shown in Table I. In the
conversion to the appropriate Z-moments, the particle
data book values for the 
hadronic branching fractions\cite{pdg} were used. 
In addition, we use Mark III results on the topological branching
fractions \cite{miii}.
The branching fractions for $D$'s into kaons is in the Particle
Data Book.
For pions, we take the branching fraction into pions 
equal to the fraction
of all decays which are {\it not} semileptonic.
The kaon multiplicities for $D^0$ and $D^+$ decays are assumed to be
$\langle N_K\rangle=1$.
To estimate $\langle N_{\pi^\pm}\rangle$, we use
the charged particle topological branching fractions. Mark III data\cite{miii}
indicate that the average charged particle multiplicities for
$D^0$ and  $D^+$ range between 2.4 and 2.6. 
We estimate that
the average {\it hadronic} 
charged particle multiplicities are between
2.1-2.4. Assuming that either $K^\pm$, $K^0$ or
$\bar{K}^0$ accompanies all charmed meson decays, and that
semileptonic decays have no pions, we estimate that the average
number of charged
pions in the decays are between 1.5-1.8, as indicated in Table I.

\section{Results}

The parameters of Table I complete what we need to evaluate the
lepton fluxes. In Figures 2 and 3, we show our resulting muon and
muon neutrino (particle plus anti-particle) fluxes scaled by a factor
of $E^3$, evaluated in the vertical direction. Our conventional,
prompt and secondary fluxes are shown by the solid lines. The dashed
lines are the TIG parameterizations which fit their Monte
Carlo results for the conventional and prompt fluxes \cite{tig}.

\vskip 0.5in
\centerline{\psfig{figure=muflux-fig2.ps,height=4.0in,angle=270}}
\vskip 0.3in
\noindent
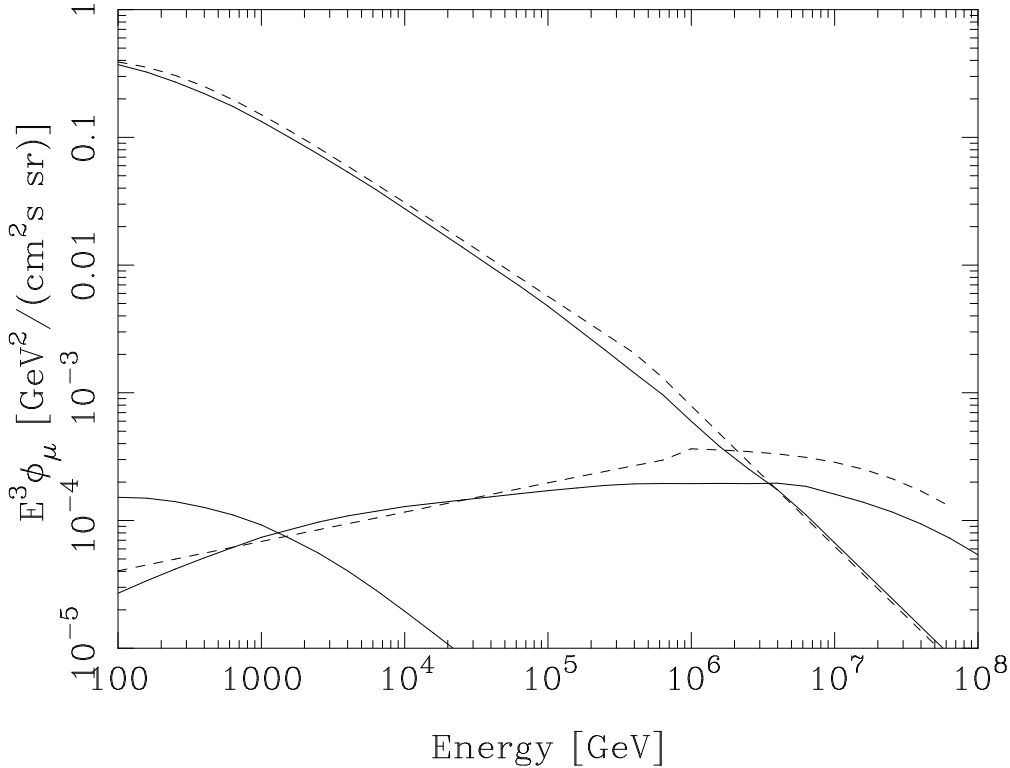
Fig. 2. The solid lines show the vertical muon plus antimuon flux,
scaled by $E^3$, as a function on muon energy $E$ for conventional,
prompt and secondary decay sources. The dashed lines show the
TIG parameterization of the conventional and prompt fluxes.
\vskip 0.3in

Before discussing the secondary fluxes, we comment on the discrepancies
between the TIG parameterization and the prompt lepton fluxes above
$E=10^5$ GeV. Part of the discrepancy comes from our ignoring
$D_s$ and $\Lambda_c$ contributions. TIG have shown that the
$\Lambda_c$, in particular, makes significant contributions at the
higher energies. In addition, while we have attempted to use the same 
parameters as TIG, not all of them are specified in their paper.
Finally, their fit was made to Monte Carlo results, which are not
exactly reproduced by this semi-analytic method using $Z$-moments.
In any case, our focus is on secondary leptons 
in an energy regime where our results
are in reasonable agreement with TIG's parameterizations.

We note that the secondary muon flux is larger than the prompt
muon flux for $E<10^3$ GeV. The secondary muon neutrino flux
exceeds the prompt neutrino flux only for $E<200$ GeV. The flux
of secondary muons (and conventional muons) is larger than the corresponding
neutrino flux because 
muons in pion decay
carry the bulk of the pion's energy. For fixed energy of the lepton,
the parent pion energy is lower for muons than for muon neutrinos.
This is a significant feature when the parent pion flux is
decreasing rapidly with energy. 

\vskip 0.5in
\centerline{\psfig{figure=nuflux-fig3.ps,height=4.0in,angle=270}}
\vskip 0.3in
\noindent
Fig. 3.
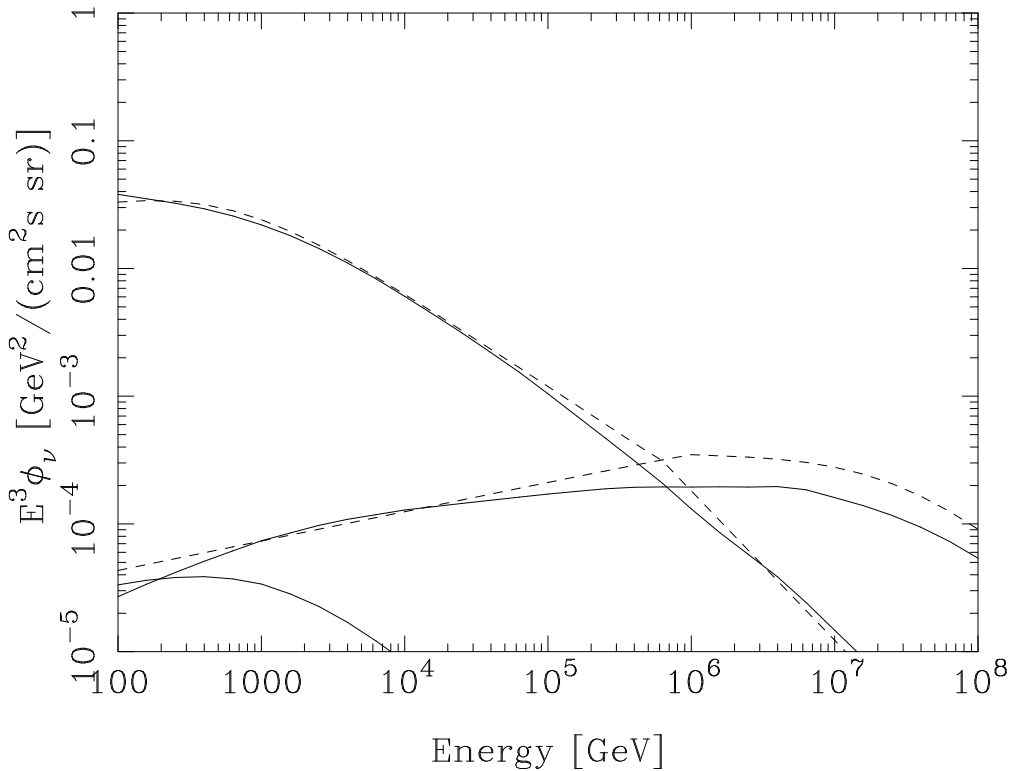
The vertical muon neutrino plus antineutrino fluxes, as in
Fig. 2.
\vskip 0.3in

The secondary
$p\rightarrow {\rm charm}\rightarrow\pi, K\rightarrow$lepton chain
is suppressed by a factor of 
\begin{equation}
R_j = {\sum_{k_c} Z_{pk_c}Z_{k_c j}\over Z_{pj} }
\end{equation}
relative to the conventional $p\rightarrow \pi,K\rightarrow$lepton chain.
For TIG's calculation of charm production based on PYTHIA,
$Z_{pk_c}\sim 10^{-3}Z_{p\pi}\sim 10^{-2}Z_{pK}$. Our secondary
fluxes are suppressed by a  factor of $\sim 1/(2.4\times 10^3)$ for muons and 
$\sim 1/(1.1\times 10^3)$ for muon
neutrinos relative to the conventional fluxes.
The errors due to our approximations in the calculation of the conventional
fluxes are more significant than the secondary flux as long as the
charm interaction and decay moments are reliably known. We turn now to the
question of whether or not the factor $R_j$ can be increased such that
the secondary flux is a significant contribution to the total flux
below lepton energies of 1 TeV.

Increasing $R_j$ may be achieved by increasing the hadronic decay
moments and the interaction moments associated with the production of
$D$'s. 
We will proceed by estimating an overall
enhancement factor $K$ with
\begin{equation}
K\equiv K_d K_\sigma K_x\ ,
\end{equation}
the product of enhancement in decay $(d)$, cross section $(\sigma )$
and in the charm energy distribution $(x=E_c/E_p)$, which may multiply
$R_j$. We'll assume that these are universal to all $R_j$.
For the limited energy range
where secondary lepton fluxes are larger than
prompt fluxes: $E=10^2-10^3$ GeV, we take
$K$ independent of energy. A full numerical study of the uncertainty
in the charm production moments as a function of energy is in
progress \cite{newprs}.

As described in Sec. III, we have evaluated the decay moments by
rescaling the $V-A$ $Z_{D\nu}$ moments according to multiplicities
and branching fractions. By using three body phase space, the decay moments
are enhanced by a factor of 2.4 at $E=10^2$ GeV. We take $K_d = 2.4$.

The cross section at next-to-leading order (NLO) in QCD is not very well
specified because of the low mass of the charm quark. A variation of
the charm quark mass between 1.2 and 1.8 GeV, and the renormalization
scale between $m_c/2$ and $2m_c$ in the NLO cross section gives a range
of cross sections that vary by more than two orders of magnitude
\cite{fmnr}. By comparing the TIG cross section with fixed target
cross section results, a factor $K_\sigma = 2$ is not unreasonable.

The $x$ distribution is an important input into the calculation of
the interaction $Z$-moments because of the steeply falling fluxes.
The quantity $x=E_c/E_p$ defined in the fixed target frame is
approximately equal to $\mid x_F\mid \equiv 2\mid p^{cm}_{||}\mid/\sqrt{s}$,
in terms of the longitudinal momentum of the charmed quark or antiquark
in the hadron center-of-momentum frame.
The theoretical (perturbative QCD) distributions in $x_F$ are typically
softer than the measured distributions, which would tend to lower the
predicted $Z$-moment. The $x_F$ distributions 
for $x_F>0$ are parameterized by
d$\sigma$/d$x_F\sim (n+1)(1-x_F)^n$.
We make the scaling assumption so that
the interaction $Z$-moments are proportional to
\begin{equation}
Z\sim \int_0^1\,{\rm d}x\, x^{1.7} A_n(1-x)^n
\end{equation}
for the proton flux proportional to $E^{-2.7}$ and normalization factor
$A_n=n+1$. For low mass charmed quarks ($m_c=1.2$ GeV), the NLO
calculations of the $x_F$ distributions, fit to a $(1-x_F)^n$ form,
give $n\simeq 6-9.5$ for $E=10^2-10^3$ GeV \cite{fmnr}. Fixed target
data yield $n\simeq 4.9-8.6 $ \cite{appel}, some with large error bars.
To estimate $K_x$, we take the NA32 value for $n$, $n=5.5$ \cite{na32}
for a typical experimental value,
and $n=7.5$ as representative of the perturbative value, so 
\begin{equation}
K_x = {\int_0^1 \,{\rm d}x\, x^{1.7} 6.5(1-x)^{5.5}\over 
\int_0^1 \,{\rm d}x\, x^{1.7} 8.5(1-x)^{7.5}} 
= 1.5\ .
\end{equation} 
Taken together, the overall enhancement factor based on modifications to
the TIG charm production parameters is on the order of $K\sim 7$.

Intrinsic charm, in principle, may boost the charm production rate.
Intrinsic charm models have a charm (and anti-charm) component
in the proton at a low scale $Q_0$, before QCD evolution has
generated $c\bar{c}$ pairs by gluon splitting \cite{brodsky}.
Estimates of the $Z$-moments for several models of intrinsic charm
were made by TIG \cite{tig}, and maximum $Z$-moments of a few times
$10^{-3}$ were found. These are at just the level of the perturbative
charm $Z$-moments, so again, may account for at most a factor of two.

Finally, we turn to a model which does not rely exclusively on the
perturbative parameters presented by TIG and energy independent modifications.
We use a charm cross section at high energies suggested by
Zas, Halzen and V\'azquez \cite{zhv}
where 
\begin{equation}
\sigma_{c\bar{c}}=0.1\sigma_
{tot}\ , 
\end{equation}
where $\sigma_{tot}$ appears in Eq. (3.2).
The form of the cross section in Eq. (4.7)
seriously overestimates the cross section in the measured regime,
$E=10^2-10^3$ GeV. We take the hadronic production cross section to be
\begin{equation}
\sigma_{c\bar{c}} = {\sigma_{c\bar{c}}^{LE}\cdot 0.1\sigma_{tot}
\over \sigma_{c\bar{c}}^{LE}+ 0.1\sigma_{tot} }
\end{equation}
where $\sigma_{c\bar{c}}^{LE}$ is the next-to-leading order perturbative
cross section with $m_c=1.3$ GeV, evaluated at the renormalization
and factorization scales set to $\mu=m_c$, for $E\leq 10^3$ GeV, using
the CTEQ3 parton distribution functions \cite{cteq}.
This is a slight overestimate of the data below $E=10^3$ GeV. Above
$10^3$ GeV, we take a power law
\begin{equation}
\sigma_{c\bar{c}}^{LE}=1.3\cdot 10^5 (E/{\rm GeV})^{0.865}\ {\rm pb}\ ,
\end{equation}
which is the power law extrapolation of the perturbative cross section 
at $E=10^3$ GeV. 

By $E\sim 10^6$ GeV, $\sigma_{c\bar{c}}\sim 0.1\sigma_{tot}$.
This assumption yields significantly higher prompt fluxes
at high energies. Indeed, Gonzalez-Garcia, Halzen, V\'azques and Zas\cite{ghv}
have shown that such large charm cross sections above 
energies of $10^5$ GeV are problematic
when one looks at Akeno horizontal air shower data\cite{akeno}.
For our purposes here, we are
interested in the cross section only as it affects the interaction
$Z$-moments below 1-10 TeV. 
For the $x$ distribution, we assume a scaling form,
d$\sigma$/d$x\sim 5(1-x)^4$, which is is the lower range of
consistency with measured values \cite{fmrc}.
Using these inputs, our results for the prompt and 
secondary muon fluxes from charm
decay are shown with the dashed lines in Fig. 4. For comparison, the solid
lines indicate the TIG results and our secondary flux calculation based
on TIG parameters, as in Fig. 2. The difference between the fluxes at
$E=10^2$ GeV are due to differences in total cross sections and
$x$-distributions relative to PYTHIA generated distributions.

\vskip 0.5in
\centerline{\psfig{figure=compare-fig4.ps,height=4.0in,angle=270}}
\vskip 0.3in
\noindent
Fig. 4.
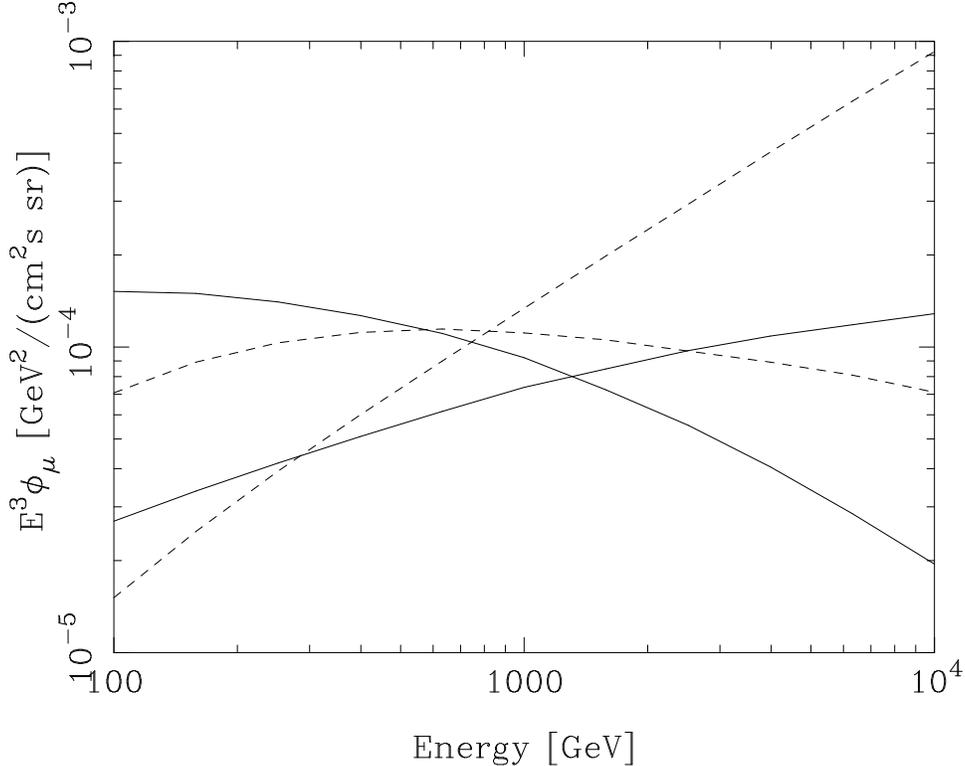
The vertical muon prompt and secondary muon fluxes, scaled by $E^3$,
using the conventional TIG parameters (solid line) and using
Eq. (4.8) and d$\sigma$/d$x\sim 5(1-x)^4$ (dashed line).
\vskip 0.3in

It is clear from Fig. 4 that at energies above a few TeV, the high
energy behavior of the cross section and $x$-distribution  have
important implications for the prompt neutrino flux. 
Recall that at these energies, the prompt muon flux equals the prompt
muon neutrino flux, and also equals the prompt electron neutrino flux.
The prompt fluxes are isotropic.
For the enhanced prompt flux
of Fig. 4, the crossover from the vertical conventional-dominated
to prompt-dominated muon neutrino
fluxes occurs around $E\sim 3\cdot 10^4$ GeV. This is in 
contrast to TIG's result of a crossover close to $E\sim 10^6$ GeV.
However, there is
no significant 
effect for $E<10^3$ GeV using the enhanced cross sections, 
given our overall uncertainty
of a factor of $K\sim 10$. 
Indeed, below $E=10^4$ GeV, a factor of 10 uncertainty
in the lepton fluxes from charm decay is a reasonable estimate.

Finally, we remark that while we have focussed on 
secondary leptons in the energy interval
100-1000 GeV, it is interesting to note that the dashed line in Fig. 4,
if valid at $E=10^4$ GeV, has important implications for the electron
neutrino flux. At $E=10^4$ GeV, the prompt muon neutrino flux is
still an order of magnitude below the vertical conventional flux, however,
the prompt electron neutrino flux is larger than the vertical
conventional flux. This can be seen by considering the ratio
$R\equiv\phi_{\nu_\mu}/\phi_{\nu_e}$ as a function of 
zenith angle $\theta$ and energy.
Using the Bartol group's calculation of the conventional atmospheric
lepton fluxes\cite{bartol}, neglecting any charm contribution, the ratios are:
\begin{eqnarray}
R_{no\ c}(E_\nu=10^4\ {\rm GeV},\cos\theta=1)& = 32\\ \nonumber
R_{no \ c}( E_\nu=10^4\ {\rm GeV},\cos\theta=0)& = 29\ .\\ \nonumber
\end{eqnarray}
However, when one adds in the prompt flux shown by the dashed line
in Fig. 4, the same ratios yield
\begin{eqnarray}
R_{with\ c}(E_\nu=10^4\ {\rm GeV},\cos\theta=1)& = 7\\ \nonumber
R_{with \ c}( E_\nu=10^4\ {\rm GeV},\cos\theta=0)& = 18\ .\\ \nonumber
\end{eqnarray}
Measuring $R$ is a difficult task, especially at 10 TeV, because of the
difficulty of measuring the electron neutrino flux. 
Air shower
arrays are, in principle, sensitive to electron neutrino 
induced events, however, acceptances
are low compared with higher energies\cite{auger}. In underground
experiments, one does not have the advantage of the long muon range
that increases the effective volume of the detector when the
process is $\nu_eN$ charged-current interactions.

\section{Conclusions}

We have focused here 
on corrections to the atmospheric lepton fluxes from charm decays in
the energy range $E=10^2-10^3$ GeV, where secondary neutrinos from
the $D\rightarrow K,\pi \rightarrow \nu_\mu,\mu$
decay chain contribute.  
The secondary flux
corrections are significant for the muon flux compared to the prompt
muon flux below $\sim 10^3$ GeV, and for the muon neutrino fluxes
below $\sim 200$ GeV. Since the branching fraction for $K^\pm$ and
$\pi^\pm$ decays to electron neutrinos is small, secondary corrections
to the atmospheric electron neutrino fluxes is negligible.

The prompt and secondary lepton fluxes are tied to the charm cross section
and energy distribution of the emerging charmed particles.
We have demonstrated here that 
the experimental constraints on the charm production cross section
and energy distribution limit the extent to which leptons from the
decays of charm can contribute to the overall atmospheric lepton fluxes.
Taking the 
parameters of TIG in Ref. [2] as a starting point, we have shown
that the uncertainties in the calculation of the lepton fluxes from
charm decay may result in an enhancement by as much as a factor of
10 in the energy range of $10^2-10^3$ GeV. 
An additional two
orders of magnitude are required to make the secondary lepton fluxes
comparable to the conventional ones at $E=100$ GeV. 
At higher energies, where one is less constrained by experiments,
the high energy behavior of the charm cross section 
and energy distribution significantly
influences the prompt flux, a topic of further investigation \cite{newprs}.

\acknowledgements
Work supported in part
by National Science Foundation Grant No.
PHY-9507688, a University of Iowa CIFRE Award and D.O.E. Contract No. 
DE-FG02-95ER40906. M.H.R. acknowledges the hospitality of the CERN
Theory Division where this work was completed. We thank P. Nason and
M. Mangano for useful discussions.

\begin{table}
\caption{Charmed particle decay moments.}
\begin{tabular}{ccccc}
Decaying Particle $(k_c)$ & $i$ & $B$ & $\langle N_i\rangle$ 
& $Z_{k_ci}(\beta_c=1.7)$\\ \tableline
 & & & & \\
$D^+$ & $K^\pm$ & 0.30 & 1 & 0.032 \\
\ & $K_L$ & 0.30 & 1 & 0.032 \\
\ & $\pi^\pm$ & 0.66 & 1.5 & 0.10 \\
\ & $\nu_\mu$,\ $\bar{\nu}_\mu$ & 0.17 & 1 & 0.018\\ 
 & & & & \\ 
$D^0$ & $K^\pm$ & 0.56 & 1 & 0.069 \\
\ & $K_L$ & 0.21 & 1 & 0.026 \\
\ & $\pi^\pm$ & 0.85 & 1.8 & 0.19 \\
\ & $\nu_\mu$,\ $\bar{\nu}_\mu$ & 0.068 & 1 & 0.0084\\ 

\end{tabular}
\end{table}

\end{document}